\documentclass[8pt,twocolumn]{article}
\usepackage[paper=a4paper, hscale=0.9, vscale=0.85]{geometry}
\usepackage{graphicx}
\usepackage[font=footnotesize]{caption}
\usepackage{sectsty}
\allsectionsfont{\normalsize\bfseries}
\usepackage{amssymb}
\usepackage{amsmath}
\usepackage{url}
\usepackage{prettyref}
	\newcommand{\pr}[1]{\prettyref{#1}}
	\newrefformat{app}{appendix~\ref{#1}}
	\newrefformat{eqn}{eqn.~(\ref{#1})}
	\newrefformat{sec}{section~\ref{#1}}
	\newrefformat{subsec}{section~\ref{#1}}
	\newrefformat{subsubsec}{section~\ref{#1}}
	\newrefformat{fig}{fig.~\ref{#1}}
\newcommand{\V}[1]{\mathbf {#1}}

\newcommand{\notes}[1]{}
\newcommand{\Rtoteg}{\gamma_{e}(N)}
\newcommand{\Rtotvirt}{\gamma_{\rm abs}(N)}
\newcommand{\lasergamma}{\beta}
\newcommand{\Fetterlambda}{\tau}
\newcommand{\SpectroscopicLO}[2]{
\raisebox{#1}{
	\includegraphics[width=#2]{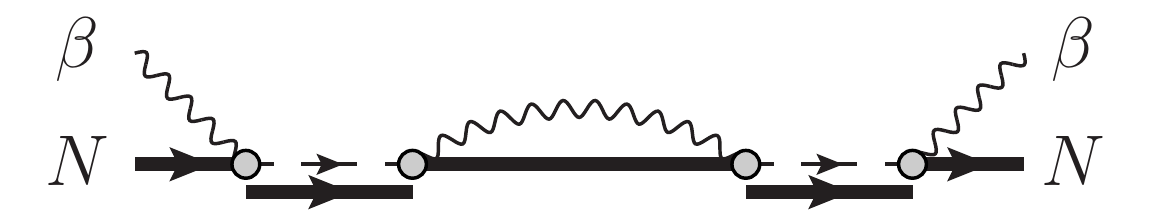}
}
}
\newcommand{\PolarizationStraightLO}[2]{
\raisebox{#1}{
	\includegraphics[width=#2]{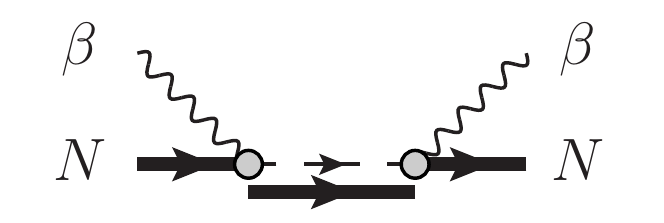}
}
}
\newcommand{\GeneralLifetime}[2]{
\raisebox{#1}{
	\includegraphics[width=#2]{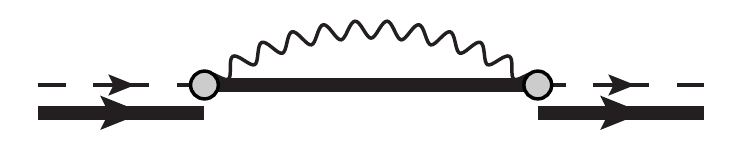}
}
}
\title{Bosonic enhancement of spontaneous emission near an interface}
\author{J\"urgen Schiefele\thanks{},  Carsten Henkel
\\
{\footnotesize Institut f\"ur Physik und Astronomie, 
Universit\"at Potsdam,
Karl-Liebknecht-Stra\ss e 24/25,
14\,476 Potsdam, 
Germany}}
\date{}
\begin{document}
\twocolumn[
\begin{@twocolumnfalse}
\maketitle
\thispagestyle{empty}
\begin{abstract}
We show how the spontaneous emission rate of an excited two-level atom 
placed in a trapped Bose-Einstein condensate of ground-state atoms is enhanced by bosonic stimulation.
This stimulation depends on the overlap of the excited matter-wave packet 
with the macroscopically occupied condensate wave function, 
and provides a probe of the spatial coherence of the Bose gas.
The effect can be used to amplify the distance-dependent decay rate 
of an excited atom near an interface.
\\
{}
\end{abstract}
\end{@twocolumnfalse}
]
{
	\renewcommand{\thefootnote}%
	{\fnsymbol{footnote}}
	\footnotetext[1]{email:\url{Juergen.Schiefele@physik.uni-potsdam.de}}
}
\section{Introduction}
\label{s:intro}
Spontaneous emission from an excited atom can only take place when a 
vacuum mode is available to accommodate
the emitted photon. 
This fact is employed in experiments in cavity quantum electrodynamics (cQED) 
to shift the spontaneous emission rate in small cavities
where the structure of the electromagnetic (em) vacuum is modified \cite{Hinds_1994, Haroche_1992}.
In particular, it is known that the lifetime of an excited atom near a plane surface 
(the simplest `cavity' system) shows an oscillatory behavior for atom-surface
distances comparable to the resonant photon wavelength. The effect is not easy
to observe with ultracold atoms since distance control in the sub-micron range
is required. 
A transient signal related to a change in decay rate was observed in the
Amsterdam group \cite{Spreeuw_2004}, with cold atoms being probed 
spectroscopically in the vicinity of a surface using an evanescent light
field.
Quite analogous to the effect of a cavity on photon modes, 
the presence of a Bose-Einstein condensate (BEC) can alter the decay 
of bosonic atoms, as the macroscopic population of atomic modes 
stimulates the transition into these \cite{Javanainen_1994,Hope_1996,Dalibard_1995}. This enhancement is
significantly reduced, however, in a uniform system because of momentum
conservation. For an excited atom initially at rest, the final state is 
shifted by the photon recoil momentum, and overlap with the 
condensate mode occurs only 
if repulsive interactions deplete the ground state \cite{Goerlitz_2001},
or alternatively, for a confined system where the BEC
is spread over a finite width in momentum space. 

In the present paper, we discuss the enhancement of spontaneous emission
in a trapped BEC and show in particular that the small oscillations in the
decay rate near an surface can be significantly amplified. The trap 
confinement and the temperature of the BEC are taken into account and provide
only a moderate reduction compared to the scaling with the number of atoms
in the BEC. We use a quantum field theory of atoms and photons to 
calculate  complex self-energies
($T$-matrix elements) \cite{Schiefele_2010}.
The spontaneous emission rate of an excited initial state is proportional to the imaginary part of
the $T$-matrix element,
while its real part yields the Lamb-shift of the initial state, and,
near a surface, the Casimir-Polder potential between BEC and surface,
as calculated in \cite{Moreno_2010, Dalvit_2008} by different methods.
We also apply the formalism to a typical 
spectroscopy experiment where the enhanced decay rate appears in the 
absorption spectrum of a weak, near-resonant laser field.
\section{Decay of an excited wavepacket in a spherical atom cloud}
\label{sec:local}
%
%
%
\subsection{Self energy and transition rate}
Consider a factorized initial state $| e, N \rangle$ with one atom in the
electronically excited state and $N$ atoms in the (collective) ground state. 
We apply second-order perturbation theory in the interaction with the em field
to get for this state a complex energy shift (self-energy) whose imaginary part 
gives the Bose-enhanced decay rate.
(This formalism is based on Refs.\cite{Lewenstein_1994, Walls_1994}
and explained in detail in Ref.\cite{Schiefele_2010}.)
The em self-energy is proportional to the $S$-matrix element
\begin{align}
\langle  e, N|
& S^{(2)} 
|N, e \rangle
        =
	\GeneralLifetime{-.5ex}{0.15\textwidth}
\label{eqn:local_Sa}
\\
	&=
	-
	\mu_\alpha \mu_\beta 
	\int d^4x_1 \, d^4x_2 \,
	\Theta(t_2-t_1)\,
	\langle \,  T \bigl\{ 
	E_\alpha(x_1) 
	E_\beta (x_2)  
	\bigr\} \, \rangle
\nonumber
\\
	&\times
	\langle e |
	\Psi_e^\dagger (x_2)  \Psi_e (x_1) 
	| e \rangle
	\langle N |
	\Psi_g (x_2) \Psi_g^\dagger (x_1)
	| N \rangle 
\label{eqn:local_S}
\; ,
\end{align}
where $\mu_\alpha$ $(\alpha = x, y, z)$ 
are the matrix elements of the transition dipole.
%
The Feynman diagram represented above uses bold lines for the 
many-body system of ground state atoms, 
dashed lines for individual excited atoms and wavy lines for the photon
propagator.
The brackets $\langle \dots \rangle$ in the first line 
of \pr{eqn:local_S} denote an expectation value with respect to 
an equilibrium state of the em-field, 
and the symbol $T{\{\dots\}}$ denotes time-ordering.
$\Psi_e$ and $\Psi_g$ are interaction-picture field operators for the atomic
levels, with a time dependence governed by the 
unperturbed atomic Hamiltonian.

In \pr{eqn:local_S},
the time dependent phase  of the ground state correlation function 
$ \langle N | \Psi_g (x_2) \Psi_g^\dagger (x_1) | N \rangle$
is of the order of typical single particle
energies in the trap and thus much smaller than the phase of the term
$\langle e | \Psi_e^\dagger (x_2)  \Psi_e (x_1) | e \rangle$,
which is proportional to 
$\operatorname{exp}[i\omega_{eg}(t_2 - t_1)]$ where
$\omega_{eg}$ is the Bohr transition frequency.
We will hence neglect the time dependence of
the ground state correlation function in \pr{eqn:local_S}. (This is
equivalent to summing the decay rate over the final ground state modes.)
The time integrations in \pr{eqn:local_S} can then be performed, 
yielding for the transition rate ($\hbar = 1$)
\begin{align}
\Rtoteg
	&=
	2 [1 + \bar n( \omega_{eg} ) ]
	\mu_\alpha \mu_\beta 
	\biggl[
	    \int d^3 x \,
	    |\Phi(\V{x})|^2
	    {\rm Im}\,{G_{\alpha\beta}(\V x,\V x,\omega_{eg})}
\nonumber
\\
	    &+
	    \int d^3 x_1 \, d^3 x_2 \,
	    \Phi(\V{x}_1) \Phi^*(\V{x}_2)\,
	    \langle N | \Psi_g^\dagger(\V{x}_1) \Psi_g(\V{x}_2) | N \rangle 
\nonumber
\\
	    &\times
	    {\rm Im}\,{G_{\alpha\beta}(\V{x}_1,\V{x}_2,\omega_{eg})}
	\biggr]
\;,
\label{eqn:local_rate}
\end{align}
where the excited state
is given by the normalized wave function $\Phi(\V x)$.
We denote $\bar n( \omega_{eg} )$ the average photon number at
frequency $\omega_{eg}$ in thermal equilibrium. 
For the sake of simplicity, we restrict ourselves for the rest of the paper to a
field at zero temperature where the thermal photon number
$\bar n( \omega_{eg} )$ is negligible.
We have expressed the photon propagator through the retarded response 
(Green) function 
\begin{align}
G_{\alpha\beta}(x_1,x_2) 
	&= 
	\int \frac{d\omega}{2 \pi}\,
	e^{i \omega (t_1 - t_2)} G_{\alpha\beta}(\V{x}_1, \V{x}_2, \omega)
\nonumber
\\
	&=
	i \langle [E_\alpha (x_1) , E_\beta(x_2)] \rangle 
	\Theta (t_1 - t_2)
\;.
\label{eqn:Gdef}
\end{align}
This quantity is easily calculated in a general environment, for example near
a surface, see Ref.\cite{Wylie_1984},
and shows oscillating behavior as function of the atom-surface separation.

In \pr{eqn:local_rate}, the decay rate naturally splits into a term
$\gamma_{e}^{(0)}$ that remains in the absence of the atom cloud
(first line)
and an additional term $\gamma_{e}^{\rm BEC}(N)$ which describes the bosonic enhancement.
The first term has a natural interpretation in terms of an average of local
decay rates over the position distribution of the excited state wavepacket.
(For a study of the dynamics of the excited state, see Ref.\cite{Japha_1998}.)
The analysis of the second term is the main
focus of this paper. Note that it depends on the two-point correlation
function of the ground state atoms. The bosonic stimulation is thus a probe
of the spatial coherence of the Bose gas.

\subsection{Discussion of Bose enhancement in free space}

To illustrate the physics in \pr{eqn:local_rate}, we will assume that both 
the ground-state BEC and the excited atom
are trapped in overlapping isotropic harmonic potentials,
far enough away from any macroscopic body. We use the free-space 
expression $G^{(0)}_{\alpha\beta}$ of the em Green tensor (\ref{eqn:Gdef}) which takes
the form
\begin{equation}
G^{(0)}_{\alpha\beta}(\V{x}_1,\V{x}_2,\omega_{eg})
	=
	\int \frac{d^3 k}{(2\pi)^3} \,
	G^{(0)}_{\alpha\beta}(k,\omega_{eg})\,
	e^{i \V k . (\V{x}_1 - \V{x}_2)}
\;.	
\end{equation}
As is well known, only photons on the light cone contribute to the 
the imaginary part of this quantity, i.e., $|{\bf k}| = \omega_{eg} / c 
\equiv k_{eg}$.
Another relevant length scale is the oscillator
length $a_0 = (M \omega_T)^{-1/2}$ of the ground-state trapping 
potential ($M$ is
the atomic mass and $\omega_T$ the trap frequency). We assume
that the excited-state wave packet $\Phi(\V x; \eta)$
is an isotropic Gaussian with a width
$\eta \, a_0$. The first term in \pr{eqn:local_rate} gives 
$\gamma_{e}^{(0)} = \mu^2 k_{eg}^3 / (3\pi\varepsilon_0)$, 
the free space decay rate \cite{Wylie_1984}.

\begin{figure}
\centering
\includegraphics[width=\columnwidth]{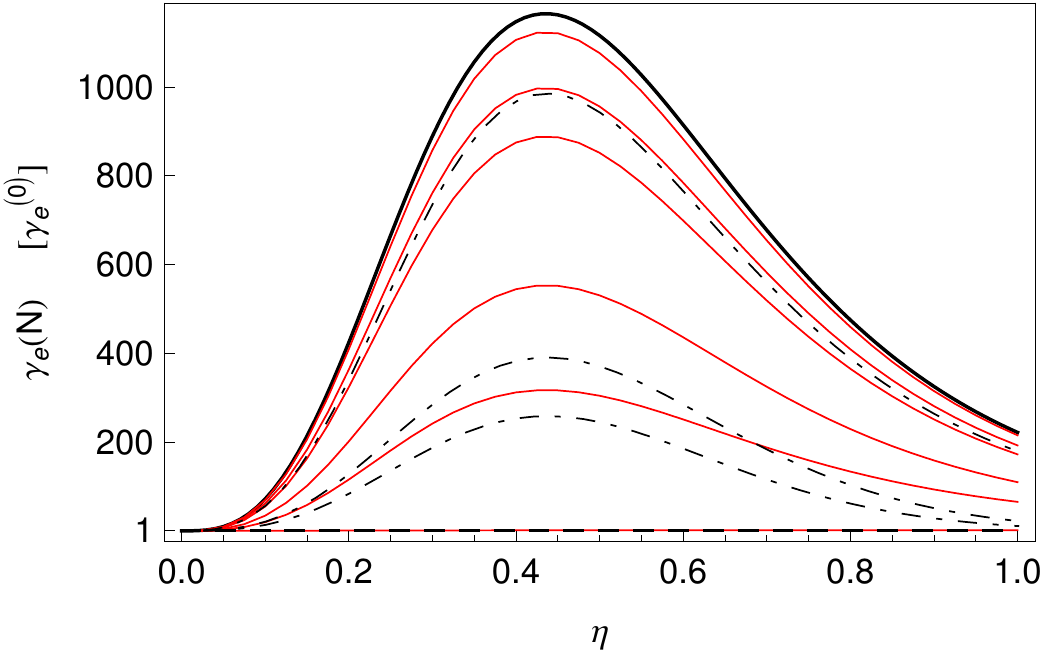}
\caption[]{Decay rate of an excited wave packet embedded in a Bose
condensate with $N$ atoms. The rate $\Rtoteg$  [\pr{eqn:local_rate}]
is normalized to the free-space value
$\gamma_{e}^{(0)} = \mu^2 k_{eg}^3 / (3\pi\varepsilon_0)$ 
and plotted 
	as a function of the width $\eta\, a_0$ of the excited state, 
	scaled to the oscillator length for the BEC trap. 
	The ground-state cloud consists of  $N = 10\,000$ atoms,
	with trap frequency  $\omega_T / 2 \pi =  1\,{\rm kHz}$
	(oscillator length $a_0 = 0.34 \,\mu{\rm m}$), the
	resonance wavelength is $2\pi / k_{eg} = 780\,{\rm nm}$ as for rubidium.
	The optimal value of $\eta$ [\pr{eqn:eta_opt}] is 
	$\eta_{\rm opt} = 0.44$.
 	\emph{Upper black curve}:
	ideal Bose gas at zero temperature.
	\emph{Solid (red) curves}:
	ideal Bose gas at temperature $T_A = 
	0.3, \, 0.5, \, 0.6, \, 0.8, \, 0.9, \, 1.2\,T_c$
	with critical temperature $T_c = \omega_T (N/ \zeta[3])^{1/3}$
	(top to bottom).
	\emph{Dash-dotted curves}:
	interacting Bose gas with mode function \pr{eqn:Fetter_psi},
	for varying s-wave scattering length $a_s = 1, \, 5, \, 
	10\, \overline {a}_s$ 	(top to bottom), with
	$k_{eg} \overline{a}_s  = 0.047$ as for rubidium.
	\emph{Horizontal dashed line}:
	free-space decay rate $\gamma_{e}^{(0)}$.
}
\label{fig:enhance_eps}
\end{figure}

Fig. \ref{fig:enhance_eps} shows how the decay rate $\Rtoteg$ varies
with the width parameter $\eta$, in the presence of a Bose condensate with
$N = 10\, 000$ Rb atoms (three-dimensional, isotropic trap).
We consider both an ideal gas model (solid red lines) and an interacting
Bose gas (dash-dotted lines). 
The upper black curve corresponds to an ideal gas at zero temperature,
where all the ground-state atoms populate the trap ground state
$\psi_{\V{0}} (\V{x})$, a gaussian with width $a_0$.
In this case the integrations in \pr{eqn:local_rate} can be worked out
explicitly. We find analytically that the optimum value of 
$\gamma_{e}( N ) $ is obtained for
%
\begin{equation}
\eta_{\rm opt}
	=
	\frac{1}{\sqrt{3}} \,
	(\sqrt{ 9 + (k_{eg} a_0)^4 } -  (k_{eg} a_0)^2 )^{1/2}
,
\quad (0 < \eta_{\rm opt} < 1)
\;.
\label{eqn:eta_opt}
\end{equation}
This result as well as the typical behaviour of $\Rtoteg$ is easy to understand 
by noting that the
decay rate for a given photon momentum ${\bf k}$ is proportional to the
overlap integral
\begin{equation}
	\int d^3x \,
	\Phi(\V x, \eta) \, \psi^*_{\V n} (\V x) \, e^{i \V k . \V x} 
\;,
\label{eqn:overlap}
\end{equation}
where ${\V n} = {\V 0}$ for the BEC ground mode. In the so-called
Lamb-Dicke limit $k_{eg} a_0 \ll 1$ (well-localized trap), the
exponential in \pr{eqn:overlap} can be approximated by unity, and the
overlap is optimal when the two wavepackets are matched in width,
$\eta = 1$. The opposite case looks closer to a homogeneous system
and is easier to analyze in Fourier space where the photon recoil
provides a shift of the momentum
distribution. This reduces the overlap and can be compensated for
by making an excited-state wave packet wider in momentum space,
i.e., $\eta < 1$. At the optimum value, the width is of the order
of the photon momentum and the shifted excited-state wave packet
still has some overlap with the sharp zero-momentum component of the BEC.


%

The temperature dependence in Fig.~\ref{fig:enhance_eps} closely
follows the occupation of the ground-state (condensate) mode.
We have used Ref.\cite{Barnett_2000} where the 
two-point correlation function 
$\langle N | \Psi_g^\dagger(\V{x}_1) \Psi_g(\V{x}_2) | N \rangle$
for the ideal Bose gas in a 3D trap 
is given in a simple form, involving only a single summation.
This provides the Bose enhancement of $\Rtoteg$ in a straightforward 
manner for atom temperatures $T_A$ below and above the critical 
temperature $T_c$ (see caption).
We see that for any $\eta$, the transition rate drops below the 
zero-temperature value, and for
temperatures above $T_c$, it becomes comparable to 
$\gamma^{(0)}_{e}$ (horizontal dashed line). 
The temperature dependence is shown in \pr{fig:spectroscopic_enhancement},
and compared to the condensate fraction $N_0 / N$ (dashed line).
The bosonic enhancement closely follows the population of the condensate 
mode because the excited-state wave packet $\Phi(\V x )$ has the 
largest overlap with the trap ground state. The thermal occupation of 
higher lying trap states hence diminishes the integrals \pr{eqn:overlap}.
We note that this behaviour would change qualitatively in lower-dimensional
systems where the Bose gas occupies excited states with a relatively 
larger weight.
%
%
%
%
%
\begin{figure}
\centering
	\includegraphics[width=\columnwidth]{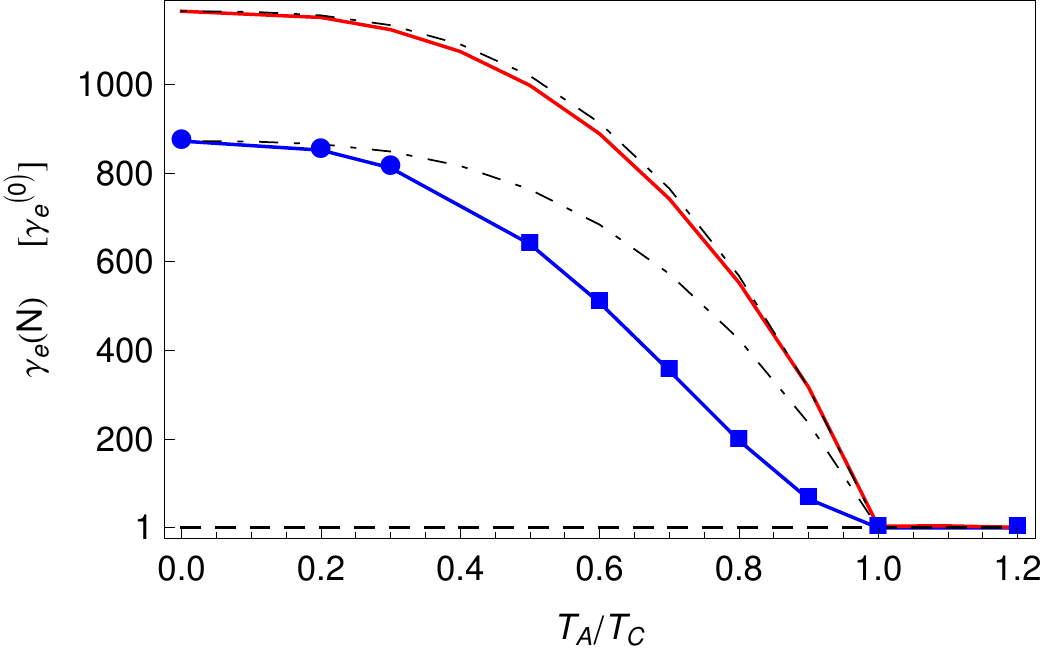}
\caption{
	Decay rate and absorption line width 
	vs temperature $T_A$ of an (ideal) Bose gas, for an optimized
	excited state wave packet. The rate $\gamma_{e}( N )$
	is given in units of the free-space value $\gamma_{e}^{(0)}$.
	Trap and atom parameters are the same as in \pr{fig:enhance_eps}.
	\emph{Red (upper) curve}: decay rate \pr{eqn:local_rate} with 
	$\eta = \eta_{\rm opt} = 0.44$.
	\emph{Blue (lower) curve}: line width $\Rtotvirt$ [\pr{eqn:spec_rate}] of the
	absorption spectrum. 
	\emph{Dashed-dotted curves}: condensate fraction $N_0/N$, scaled to
	the $T = 0$ values.
	\emph{Horizontal dashed line}:
	free-space decay rate $\gamma_{e}^{(0)}$.
	}
\label{fig:spectroscopic_enhancement}
\end{figure}
%

The more realistic case of an interacting Bose gas is also shown in
Fig.\ref{fig:enhance_eps}. We focus here on repulsive interactions
(corresponding to positive s-wave scattering length $a_s$),
and restrict ourselves to temperatures far below $T_c$, where it 
is legitimate to approximate the field operator
by the condensate mode only:
\begin{equation}
\langle N | \Psi_g^\dagger(\V{x}_1) \Psi_g(\V{x}_2) | N \rangle
	\approx
	N \psi^*_{\V{0}} (\V{x}_1) \psi_{\V{0}} (\V{x}_2)
\;.
\label{eqn:corr_app}
\end{equation}
%
Elementary excitations of the condensate can be included within Bogoliubov 
theory~\cite{Stringari_1996, Hu_2004, Oehberg_1997}.
The condensate wave function $\psi_{\V{0}} (\V{x})$ is a solution of the Gross-Pitaevskii equation.
We have used the approximate variational solution \cite{Fetter_1997}
\begin{equation}
\psi_{\V 0}(\V x)
	=
	\frac{c_0( \Fetterlambda )}{N_0 R^{3/2}} \,
	(1 - x^2/R^2)^{(1+\Fetterlambda)/2} \,
	\Theta(R - x)
\;,
\label{eqn:Fetter_psi}
\end{equation}
\notes{TF_enhancement.nb}
that interpolates between a Gaussian and the Thomas-Fermi limit as the
parameter $N_0 a_s / a_0$ is changing from zero to infinity. The length
$R$ and the exponent $\Fetterlambda$ are fixed by minimizing the Gross-Pitaevskii
energy functional, and $c_0$ is a normalization constant.
The result for the decay rate $\Rtoteg$ is shown by the dashed-dotted curves 
in \pr{fig:enhance_eps}, 
as the interaction parameter $N_0 a_s / a_0$ is increased. Relative
to the s-wave scattering length $\overline {a}_s$ of rubidium, we took
$a_s = 1, \, 5, \, 10\, \overline {a}_s$ (top to bottom) which can be 
achieved using a Feshbach resonance, for example. The interacting gas
shows a flatter density profile in the trap, as is well known; this results in
smaller values of the overlap integrals~\pr{eqn:overlap}.

To summarize the data of Fig.\ref{fig:enhance_eps}, we find a relatively
strong enhancement of the spontaneous decay rate of an excited atom
embedded in a Bose condensate. This happens despite the non-perfect
overlap that encodes the constraints of momentum conservation and
photon recoil. The optimum conditions correspond to a well-localized 
excited-state wavepacket (on the scale of the transition wavelength)
and a strong condensate fraction ($T \lesssim 0.5\, T_c$).

%
%
\section{Bose enhancement near a surface}
\label{sec:near-a-surface}
\notes{math/pancake30.nb}
In this section, we calculate the transition rate \pr{eqn:local_rate} near
a surface and demonstrate its enhancement in a Bose condensate of oblate 
shape.
This scenario can be realized with an optical lattice, by retro-reflecting
an off-resonant laser beam at the surface \cite{Spreeuw_1995},
or in a bichromatic evanescent wave \cite{Ovchinnikov_1991}. We take 
the surface in the $xy$-plane and the trapped atoms centered at 
a distance $d$ from the surface in the positive $z$-direction. 
Concerning the surface material, we use the idealized model of a perfectly 
reflecting mirror for the sake of simplicity;
but with the appropriate choice of (frequency-dependent) reflection 
coefficients that appear in the photonic Green function 
$G_{\alpha\beta}( {\rm x}_1, {\rm x}_2, \omega )$, 
a wide range 
of surface materials can be treated in the same manner\cite{Sipe_1981,Wylie_1984}.

As we have measurements in mind where the control over the distance $d$ 
is essential, we take an oblate Bose condensate 
and assume for simplicity a single gaussian mode with widths
$a_0$ (in the $xy$-plane) and $a_0 / \sqrt{\lambda} \ll d$ (along the $z$-axis).
The depletion of the condensate and its broadening due to repulsive 
interactions could be incorporated as in Sec.\ref{sec:local}.
For the excited state, we adopt again a gaussian wave packet
localized in the cloud center, with widths $\eta a_0$ and
$\eta_z a_0 / \sqrt{\lambda}$, respectively. The actual values of the trap
frequency are given in the caption of \pr{fig:loose}.
As the very narrow confinement in the $z$-direction describes a quasi-2D scenario,
the temperature has to be lower than $T_c^{(2D)} = \omega_\parallel (N/\zeta[2])^{1/2}$ to ensure a strong condensate occupation.
%
%
\begin{figure}[t]
\centering
	\includegraphics[width=\columnwidth]{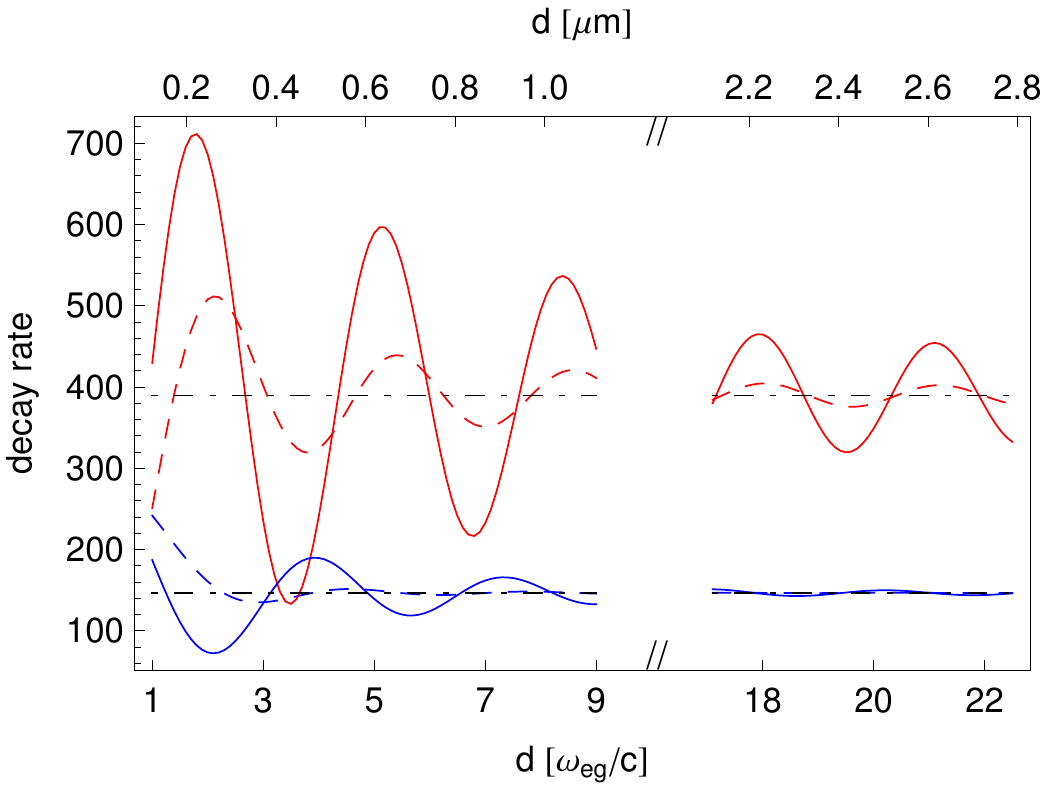}
\caption{Bose-enhanced decay rate $\Rtoteg$ near an interface.
	The BEC contains $10^5$ Rb atoms in an oblate wave function
	at an average distance $d$ from a perfectly reflecting surface.
	The size parameters are $a_0 = 3.4 \,\mu{\rm m}$ parallel and
	$a_0/\sqrt{\lambda} = 0.01\, a_0$ perpendicular to the surface,
	corresponding to trapping frequencies of $\omega_{\parallel} / 2 \pi= 10\,{\rm Hz}$ 
	and $\omega_\perp = 10^4 \omega_\parallel$ and 
	a critical temperature $T_c^{(2D)} = 118\,{\rm nK}$.
	The excited wave packet 
	(resonance frequency as in \pr{fig:enhance_eps}) 
	is spatially centered in the BEC, with size parameters 
	$\eta = 0.07$ and $\eta_z  = 1$ relative to $a_0$.
	\emph{Full red curve}: Bose-enhanced decay rate
	$\gamma_{e}^{\rm BEC}( N )$ given by \pr{eqn:local_rate}, 
	for an excited atom with its dipole moment oriented parallel to the surface.
	We normalize to
	the free-space decay rate $\gamma_{e}^{(0)}( d \to \infty )$.
	\emph{Full blue curve}: 
	$\gamma_{e}^{\rm BEC}( N )$ 
	for an excited atom with perpendicular dipole moment. 
	\emph{Horizontal dashed lines}:
	asymptotic values of $\gamma_{e}^{\rm BEC}( N, d \to \infty )$
	at large separation.
	\emph{Dashed red (blue) curve}: single-atom decay rate
	$\gamma_{e}^{(0)}$ (first line of
	\pr{eqn:local_rate}) for parallel (perpendicular) dipole orientation;
	these data are 
	multiplied by a factor of 390 and 150, respectively,
	such that their asymptotic values for large distances $d$ coincide with
	$\gamma_{e}^{\rm BEC}( N )$.
} 
\label{fig:loose}
\end{figure}

Fig. \ref{fig:loose} illustrates the decay rate $\Rtoteg$ obtained from
\pr{eqn:local_rate} 
as a function of the distance $d$. As is well known, the rate 
depends on the orientation
of the dipole moment (parallel or perpendicular to the surface, represented
in red and blue, respectively). 
The full curves show the Bose-enhanced contribution
$\gamma_{e}^{\rm BEC}( N, d )$, while 
the dashed curves give the single-atom part $\gamma_{e}^{(0)}( d )$,
re-scaled such that the asymptotic value for large distances $d$ coincides 
with $\gamma_{e}^{\rm BEC}( N, d \to \infty )$.
The numbers given in \pr{fig:loose} are the result of a compromise
between a tight confinement in the vertical ($z$-) direction and
a localized wave packet in the excited state. 
The atomic wave packets must be confined below the wavelength in
the $z$-direction, otherwise the oscillations in $\gamma_{e}$ vs.\
distance are averaged out. In this limit, the optimal Bose enhancement
is found for an excited wave packet that is matched to the condensate
($\eta_z = 1$). For the size parameter in the $xy$-plane,
we find an optimum at $\eta = 0.07$.
The asymptotic values $\gamma_{e}(N, {d \to \infty})$ are enhanced by 
factors of 
$390$ and $150$ compared to $\gamma_{e}^{(0)}( {d \to \infty} )$ for the
parallel and perpendicular dipole, respectively. The difference between these
two numbers and the relative phase shift of the oscillation pattern
in $\gamma_{e}^{\rm BEC}( d )$ are due to the radiation pattern
of the dipole emission, combined with the shape of the ground state mode
that modulates the Bose enhancement in ${\bf k}$-space.
%

Fig. \ref{fig:loose} thus
demonstrates a significant amplification of the decay rate above the surface,
with the oscillation amplitude receiving an additional enhancement relative to the  asymptotic
free-space component.
It suggests that even at a distance
of a few microns (several transition wavelengths), Bose enhancement
can bring the tiny interference structure of the decay rate into an
experimentally detectable regime.
%
%
%
\section{Virtual excited atoms produced by laser absorption}
The calculation above assumed the presence of an excited atom prepared 
in a gaussian wave packet, and one may ask the question whether this is
a realistic description. Indeed, the preparation of such a state would typically
proceed by illuminating the system. We therefore describe in this section
a calculation of a typical absorption spectrum. We find that the results of
the previous section are qualitatively unchanged. The method also illustrates
the relevance of two- and four-point correlation functions of the Bose gas.
For the sake of simplicity, we restrict this analysis to the ideal Bose gas.

The calculation proceeds by keeping a continuum of modes for the excited
state field operator $\Psi_e( x )$ and by identifying the absorption
spectrum of a weak laser field with a suitable $T$-matrix element (self-energy).
We take the laser field to be described by a coherent state 
$| \lasergamma \rangle$ in
a given plane-wave mode.

In the leading order of perturbation theory,
the absorption by the atom cloud of a photon out of the coherent state 
$|\lasergamma\rangle$ 
and re-emitting it into the same state,
\begin{equation}
\PolarizationStraightLO{-1.3ex}{0.30\columnwidth}
\;,
\end{equation}
results in a (complex) energy shift 
of the laser plus atom system that is described by 
the $T$-matrix element
\begin{equation}
\langle N, \lasergamma | T^{(2)} | N, \lasergamma \rangle
	=
	\frac{|\lasergamma|^2 \omega_L}{2} 
	\hat{e}_\alpha(\V{k}_L) 
	\hat{e}_\beta(\V{k}_L) 
	\frac{N \mu_\alpha  \mu_\beta}{\omega_{eg} - \omega_L - i \epsilon } \,
\;.	
\label{eqn:alpha_LO_1}
\end{equation}
%
In \pr{eqn:alpha_LO_1}, $|\lasergamma|^2$ is the number of photons in the coherent state, 
$\omega_L$ and $\V{k}_L$ denote the frequency and wave vector of the absorbed photons, 
the unit vectors $\hat{\V e}(\V{k}_L)$  denote axes of (linear) photon  polarization,  
and the infinitesimal $\epsilon \searrow 0$ ensures the adiabatic switching-on
of the laser field.
At this order of perturbation theory, the absorption of photons by the atom 
cloud is proportional to
$
{\rm Im}\,{\langle N, \lasergamma | T^{(2)} | N, \lasergamma \rangle} 
\propto
\delta(\omega_{eg} - \omega_{L}).
$

The next order in perturbation theory brings about the diagram
\begin{equation}
	\SpectroscopicLO{-1.1ex}{0.45\columnwidth}
\;,
\label{eqn:graph_NLO}
\end{equation}
which gives the following contribution to the $T$-matrix:
%
\begin{align}
\langle N, \lasergamma |&
T^{(4)} | N, \lasergamma \rangle
	=
	-\frac{|\lasergamma|^2 \omega_L}{2} 
	\hat{e}_\alpha(\V{k}_L) 
	\hat{e}_\beta(\V{k}_L) 
	\frac{ \mu_\alpha \mu_\beta }{[\omega_{eg} - \omega_L - i \epsilon ]^2}
\nonumber
\\	
	&\times
	\int d^3 {x}_1 \int d^3 {x}_2 \,
	\langle \Psi_g^\dagger (\V {x}_2) \Psi_g (\V {x}_2) \Psi_g^\dagger (\V {x}_1) \Psi_g (\V {x}_1)\rangle \,
\nonumber
\\	
	&\times
	\mu_\gamma  \mu_\delta 
	G_{\gamma\delta} (\V {x}_1, \V {x}_2, \omega_{L})\,
	e^{-i \V{k}_L . (\V {x}_1 - \V {x}_2)}
\;.
\label{eqn:T_NLO}
\end{align}
Let us introduce the density correlation function of the Bose gas as
\begin{equation}
	C( \V {x}_2,  \V {x}_1 )  =
	\langle \Psi_g^\dagger (\V {x}_2) \Psi_g (\V {x}_2) \Psi_g^\dagger (\V {x}_1) \Psi_g (\V {x}_1)\rangle 
	- n(\V {x}_2) 
	n(\V {x}_1)
	\label{eq:def-density-correlations}
\end{equation}
where $n(\V {x}_1) = 
\langle \Psi_g^\dagger (\V {x}) \Psi_g (\V {x})\rangle$ is the average
density.
%
This splits \pr{eqn:T_NLO} in two parts:
$
\langle  T^{(4)}  \rangle
	=
	\langle  T^{(4)}_{\rm scat}  \rangle
	+
	\langle  T^{(4)}_{\rm abs}  \rangle
.
$
The former contains only densities and can be identified with the
elastic scattering of photons off the inhomogeneous density profile 
of the BEC.
This term does not distinguish between
a Bose gas and a classical system with the same density. Its imaginary
part provides, by the optical theorem, the total scattering cross section
of the BEC.
The second term $\langle  T^{(4)}_{\rm abs}  \rangle$, on the contrary,
is proportional to density fluctuations, and these are at the origin of 
bosonic enhancement \cite{Ketterle_2001,Moore_2001}. We therefore
identify ${\rm Im}\,{\langle  T^{(4)}_{\rm abs}  \rangle}$ with the change in 
the atomic absorption spectrum (line width).

Indeed, if we define the resonant part of the polarizability 
$\alpha_{\alpha\beta}(\omega)$ of the atom cloud as
\begin{equation}
\alpha^{\rm res}_{\alpha\beta}(\omega)
	=
	\frac{N \mu_\alpha  \mu_\beta}{\omega_{eg} - \omega - i \epsilon }
\;.
\label{eqn:alpha_LO_2}
\end{equation}
we see that the process \pr{eqn:graph_NLO} can be re-written as
a shift of the atomic transition frequency
$\omega_{eg} \to \omega_{eg} + \delta \omega_{eg}$ with 
\begin{align}
\langle  T^{(2)}  \rangle + \langle  T^{(4)}_{\rm abs}  \rangle
	&=
	\frac{|\beta|^2 \omega_L}{2} 
	\hat{e}_\alpha(\V{k}_L) 
	\hat{e}_\beta(\V{k}_L) 
\nonumber
\\	
	&\times
	\bigl[
	\alpha^{\rm res}_{\alpha\beta}(\omega_L) 
	+
	\delta \omega_{eg} \,
	\frac{\partial \alpha^{\rm res}_{\alpha\beta}(\omega_L)}{\partial \omega_{eg}}
	\bigr]
\;.
\label{eqn:model-for-delta-alpha}
\end{align}
By identifying \pr{eqn:T_NLO} and \pr{eqn:model-for-delta-alpha},
we can read off the frequency shift $\delta \omega_{eg}$ whose
imaginary part yields the atomic line width (the inverse lifetime of the
virtual state involving an excited atom)
%
\begin{align}
&\Rtotvirt
	=
	- 2 \, {\rm Im}\,{\delta \omega_{eg}} 
\nonumber
\\
	&=
	\frac 2 N
	\mu_\alpha 
	\mu_\beta
	\biggl[
	\int d^3 {x} \,
	n( \V x) \;
	{\rm Im}\,{ 
	G_{\alpha\beta} (\V x, \V x, \omega_{L})
	}
\nonumber
\\
	&+
	\int d^3 {x}_1 \int d^3 {x}_2 
	C( {\bf x}_1, {\bf x}_2 ) 
	{\rm Im}\,{ 
	G_{\alpha\beta} (\V {x}_1, \V {x}_2, \omega_{L})  
	e^{-i \V{k}_L . (\V {x}_1 - \V {x}_2)}
	}
	\biggr]
.
\label{eqn:spec_rate}
\end{align}
This function depends weakly on the laser frequency, and we evaluate
it at $\omega_L = \omega_{eg}$ for simplicity. We shall use below
Wick's theorem to evaluate the density correlation function
\cite{Evans_1996}, as appropriate for the ideal Bose gas:
\begin{equation}
	C( {\bf x}_1, {\bf x}_2 ) =
	\left| \langle \Psi_g^\dagger (\V {x}_2) \Psi_g (\V {x}_1) \rangle 
	\right|^2
\label{eqn:Wick-for-density-correlation}
\end{equation}
Eqs.(\ref{eqn:spec_rate}, \ref{eqn:Wick-for-density-correlation}) 
can now be compared to the 
decay rate $\Rtoteg$ of an excited-state wave packet~(\pr{eqn:local_rate}).
The first term in both expressions is very similar, and we see that the
laser spectroscopy effectively prepares an excited state density profile
matched to the condensate density. The second terms differ because the
laser wave-vector appears explicitly. Also
the one-body density matrix for the excited state,
$\langle e | \Psi_e^\dagger (x_2)  \Psi_e (x_1) | e \rangle$
in \pr{eqn:local_S}, is replaced by its ground-state equivalent
$\langle N | \Psi_g^\dagger (\V {x}_2) \Psi_g (\V {x}_1) | N \rangle$ 
in \pr{eqn:spec_rate}. The prepared wavepacket is hence no longer pure.
This makes the temperature dependence of $\Rtotvirt$ stronger, as can
be seen in Fig.\ref{fig:spectroscopic_enhancement} (compare the
blue and red curves). The calculation of
$\Rtotvirt$ involves, because of the squared correlation function,
a double summation over single-particle trap states \cite{Barnett_2000}
under the integral. 
At zero temperature, the summations can be done analytically, at the
low temperatures $T_A = 0.2, \, 0.3 \, T_c$, the double sum could be
evaluated numerically (circles in \pr{fig:spectroscopic_enhancement}),
while for $T_A/T_c \geq 0.5$, the summations can be accurately replaced
by integrations that evaluate faster (denoted by squares).
The size parameter $\eta$ was set to the optimal value obtained from
Fig.\ref{fig:enhance_eps}.
Although the line width $\Rtotvirt$ is for these parameters around 30\% 
smaller than the optimized decay rate $\Rtoteg$, the strong Bose
enhancement is still working in a qualitatively similar way for both types
of processes. We expect a similar result to hold for an absorption experiment
near a surface, using for example evanescent fields as discussed in 
Refs.\cite{Courtois_1995, Aspect_1995, Spreeuw_2002}.
%
%
%
%
%
%
%
%
\section{Summary}
To summarize, the presence of a trapped BEC can significantly enhance
the decay of an excited atom by bosonic stimulation. 
The magnitude of the effect depends on the overlap between the atomic
wave functions and the wavevector of the photon involved in the decay.
More precisely, our calculations based on a quantum field theory of the
atom-photon interaction illustrate the importance of two- and four-point 
correlation functions of the ground-state field for the Bose enhancement.
For an excited atom prepared in a Gaussian wavepacket, the transition rate 
to the ground state can be increased under optimum conditions by a 
factor $N/10$ where $N$ is the atom number in the BEC.
This effect also amplifies the small oscillations of the decay rate 
near an interface. We have provided an alternative calculation based on
the absorption of a laser beam that  qualitatively confirms the simpler
wave packet picture. The main difference is that absorption from the laser
field prepares a non-pure excited state which matches the one-body density
matrix of the Bose gas.
%
%
%
%
%
%
\section*{Acknowledgement}
This work was supported by \emph{Deut\-sche For\-schungs\-ge\-mein\-schaft} (DFG).
We thank Dalimil Maz\'{a}\v{c} for fruitful discussions.
\bibliographystyle{elsarticle-num}
{\footnotesize
\bibliography{bose-cQED}
}
\end{document}